\documentclass[]{pasj02} 
\usepackage[switch,mathlines]{lineno} 

\jyear{2026}
\Received{}
\Accepted{}

\usepackage{comment}
\usepackage{natbib}
\usepackage{enumitem}
\usepackage{color}

\begin{document} 

\title{Demographic dynamics of the Japanese astronomy and astrophysics publication community during 1996--2025}

\author{Hideaki \textsc{Fujiwara}\orcid{0000-0001-6536-8656} }
\affil{Frontier Research Institute for Interdisciplinary Sciences (FRIS), Tohoku University, 6-3 Aramaki-aza Aoba, Aoba-ku, Sendai, Miyagi 980-8578, Japan}
\email{hideaki@fris.tohoku.ac.jp}


\maketitle

\begin{abstract}
We investigate the demographic evolution of the Japanese astronomy and astrophysics publication community during 1996--2025 using bibliometric data from 23,274 publications in major journals in the field. The number of active Japan-affiliated authors increased from approximately 600 in the late 1990s to more than 2,000 by the early 2020s, while annual publication output reached approximately 1,400 by 2025. Participation was highly heterogeneous, with 46\% of authors observed in only one publication year within the study period and a smaller core showing long observed participation periods. We define the Community Renewal Rate (CRR) as a community-level measure of renewal, calculated as the fraction of active authors not observed during a specified preceding interval. The 5-year CRR typically remained near one quarter of the active community, with annual values ranging from approximately 21\% to 33\%, indicating persistent renewal despite substantial community growth. An order-of-magnitude comparison suggests that newly entering students account for only part of the renewal-author population. The community therefore expanded while sustaining both continuity through authors with long observed participation periods and renewal through recurring changes in active membership.
\end{abstract}


\section{Introduction}\label{sec:1}

Scientific communities evolve continuously as researchers join, leave, and re-engage over time. Understanding these demographic dynamics is important for characterizing the long-term development and sustainability of scientific disciplines. However, the composition and renewal of research communities have received less attention than research production, citation impact, and collaboration.

Official statistics describe the size and composition of the research workforce through institutional surveys. In Japan, researcher numbers are derived primarily from the national Survey of Research and Development \citep{RD2025}, a questionnaire-based survey covering business enterprises, public and nonprofit research institutions, and universities and colleges, using definitions of research personnel broadly consistent with the OECD Frascati Manual \citep{OECDFrascatiManual, NISTEP2025}. Such statistics provide essential information for science and technology policy, but they characterize an institutionally defined research workforce rather than directly identifying researchers who are actively contributing to the scientific literature.

Publication databases provide a complementary perspective by identifying active research communities through scholarly outputs and publication-time affiliations. Longitudinal bibliometric data have been used to examine publishing careers, attrition, international mobility, and return migration \citep{Kwiek2025,RobinsonGarcia2019,Zhao2022}. These studies primarily focus on individual researchers or cohorts, with relatively little attention paid to publication-community renewal, including both first-time participants and authors returning after prolonged publication gaps.

In astronomy and astrophysics, bibliometric studies have characterized journals, facilities, collaborations, and national or institutional research profiles \citep[e.g.,][]{Abt1981,Abt1996,Moed2010,Bornmann2012,Waltman2016}. Recent bibliometric studies have examined the scholarly impact of large astronomical facilities \citep{Fujiwara2025}, benchmarked astronomy journals \citep{Fujiwara2026a}, and characterized the institutional evolution of the Japanese astronomy and astrophysics research community \citep{Fujiwara2026b}.

In this paper, we investigate the long-term demographic dynamics of the Japanese astronomy and astrophysics publication community during 1996--2025 using longitudinal bibliometric data. We characterize its evolution from the perspectives of community size, participation duration, community renewal, and author turnover. To quantify community renewal, we define the Community Renewal Rate (CRR) as the fraction of active authors who had no observed participation in the community during a specified preceding interval. The CRR captures both first-time participants and authors returning after prolonged publication gaps, thereby distinguishing community renewal from first-time entry alone.

\section{Data and methodology}\label{sec:2}
\subsection{Dataset construction}\label{ssec:21}

We used Elsevier's Scopus\footnote{$\langle$https://www.scopus.com/$\rangle$.} as the primary bibliographic data source because it provides persistent author identifiers and publication-time affiliation metadata suitable for constructing a longitudinal author registry. Although the NASA Astrophysics Data System (ADS)\footnote{$\langle$https://ui.adsabs.harvard.edu/$\rangle$.} is the standard bibliographic database in astronomy, it does not provide these data in the required form.

The initial publication list was obtained from Elsevier's research analytics platform SciVal\footnote{$\langle$https://www.scival.com/$\rangle$.}, which is built on Scopus data. We first identified publications classified within the subject area ``Astronomy and Astrophysics'' (All Science Journal Classifications (ASJC) code 3103) over the year range 1996--2025 and then restricted the results to publications associated with Japan in the SciVal country/region filter. We further limited the dataset to articles, reviews, and letters in \textit{Astronomy \& Astrophysics}, \textit{The Astronomical Journal}, \textit{The Astrophysical Journal}, \textit{The Astrophysical Journal Letters}, \textit{The Astrophysical Journal Supplement Series}, \textit{Monthly Notices of the Royal Astronomical Society}, \textit{Monthly Notices of the Royal Astronomical Society: Letters}, \textit{Nature Astronomy}, \textit{Publications of the Astronomical Society of Japan}, and \textit{Publications of the Astronomical Society of the Pacific}. This journal set follows the longitudinal bibliometric framework established in our previous studies \citep{Fujiwara2026a} and was selected to capture the principal international and domestic publication venues of Japanese astronomy and astrophysics. The resulting publication list contained 23,274 publications. The underlying Scopus database was last updated on July 1, 2026, and the publication list was exported on July 10, 2026. The SciVal country/region filter was used only to construct the initial publication list. Japan-affiliated authors were subsequently identified independently from the full Scopus records using the conservative publication-time affiliation criteria described below.

From the SciVal export, we used only publication years and Scopus electronic identifiers (EIDs). Author and affiliation fields from the export were not used because long lists may be truncated and complete author--affiliation relationships are not preserved. Complete metadata for all 23,274 target publications were therefore retrieved from the Scopus Abstract Retrieval Application Programming Interface (API) using their EIDs. The returned full records were retained for reconstruction of author--affiliation relationships.

For each publication, publication-time affiliations were reconstructed from the Scopus author-group structure. We adopted a conservative rule to identify affiliations located in Japan. An affiliation was classified as Japanese when its structured country metadata explicitly indicated Japan, either through the country name \texttt{Japan} or the country code \texttt{jpn}. When structured country metadata were absent, an affiliation was also classified as Japanese if affiliation-related text within the same author-group record, including organization, source-text, or city fields, explicitly contained the word \texttt{Japan}. In contrast, when the structured country metadata explicitly indicated another country, that classification was retained even if the affiliation text contained the word \texttt{Japan}. 

An author was classified as Japan-affiliated for a given publication when at least one of the author's publication-time affiliations satisfied the above criterion. Authors without a Scopus Author Identifier (ID) were excluded. The Scopus Author ID was then used to link records belonging to the same individual across publications and years. The resulting author--publication--affiliation records were deduplicated to construct an annual registry of Japan-affiliated authors. For each calendar year, the active publication community was defined as the set of unique Scopus Author IDs associated with at least one Japan-affiliated publication-time affiliation in at least one selected publication during that year. An author appearing in multiple publications or with multiple Japanese affiliations in the same year was therefore counted only once.

First-author activity was examined as a complementary indicator of author-position patterns within the expanding publication community. The annual active-author count measures participation in the selected literature irrespective of author position, while first-author activity provides a simple and consistently available measure of participation in the first-listed position of the published author sequence. For these analyses, the first author of each publication was defined operationally as the first-listed author in the Scopus author sequence. A publication was classified as having a Japan-affiliated first author when this first-listed author satisfied the same publication-time affiliation criterion described above, and annual counts of first authors were based on unique Scopus Author IDs. Because this definition is based solely on author order, it does not capture co-first authors or equal-contribution designations that may be indicated in the full text or publisher metadata. The first-author measures should therefore be interpreted as first-listed-author indicators rather than as complete measures of authorship contribution or responsibility.

Under this procedure, 23,269 of the 23,274 target publications contained at least one identifiable Japan-affiliated author, corresponding to a publication-level coverage of 99.98\%. The resulting registry contained 9,910 unique Japan-affiliated authors over the full 1996--2025 period. The dataset therefore represents the astronomy and astrophysics publication community associated with institutions located in Japan at the time of publication, rather than the entire astronomy research workforce in Japan.

\subsection{Demographic indicators}\label{ssec:22}

To characterize the demographic evolution of the publication community, we consider the Community Renewal Rate (CRR), the entry rate, and the exit rate. Among these, the CRR is the primary indicator used to characterize community renewal in this study, while the entry and exit rates are used only to facilitate its interpretation.

For a given year $Y$, let $N(Y)$ denote the number of active authors in the publication community. We define a \textit{renewal author} as an author who published in year $Y$ but had no publication associated with the community during the preceding $T$ years. If the number of renewal authors is denoted by $R_T(Y)$, the CRR is defined as
\begin{equation}
\mathrm{CRR}_T(Y)=\frac{R_T(Y)}{N(Y)}.
\end{equation}

Renewal authors include both first observed participants and authors returning after prolonged publication gaps, as well as authors entering from overseas or other fields. The CRR therefore measures renewed active membership rather than first-time inflow alone. In this study, $T=1$, 3, 5, and 10 years were adopted to characterize community renewal over different timescales.

For comparison, we also calculated the entry rate and exit rate. The entry rate is defined as the fraction of authors active in year $Y$ whose first observed publication in the community occurred in year $Y$. Conversely, the exit rate is defined as the fraction of authors active in year $Y$ whose last observed publication in the community occurred in year $Y$. The entry rate approximates the influx of authors into the publication community, while the exit rate approximates their departure from, or long-term inactivity within, the community. Both indicators are affected by the finite temporal coverage of the dataset, unlike the CRR. The entry rate is subject to left-censoring because publication records before 1996 are unavailable, while the exit rate is subject to right-censoring because publications after 2025 are not included in the present analysis. These two indicators are therefore used only as supplementary measures for interpreting the CRR. The 5-year CRR is reported only from 2001 onward because its evaluation requires a complete 5-year observation history.

\section{Results}\label{sec:3}

\subsection{Community size and publication activity}\label{ssec:31}

\begin{figure}
 \begin{center}
  \includegraphics[width=90mm]{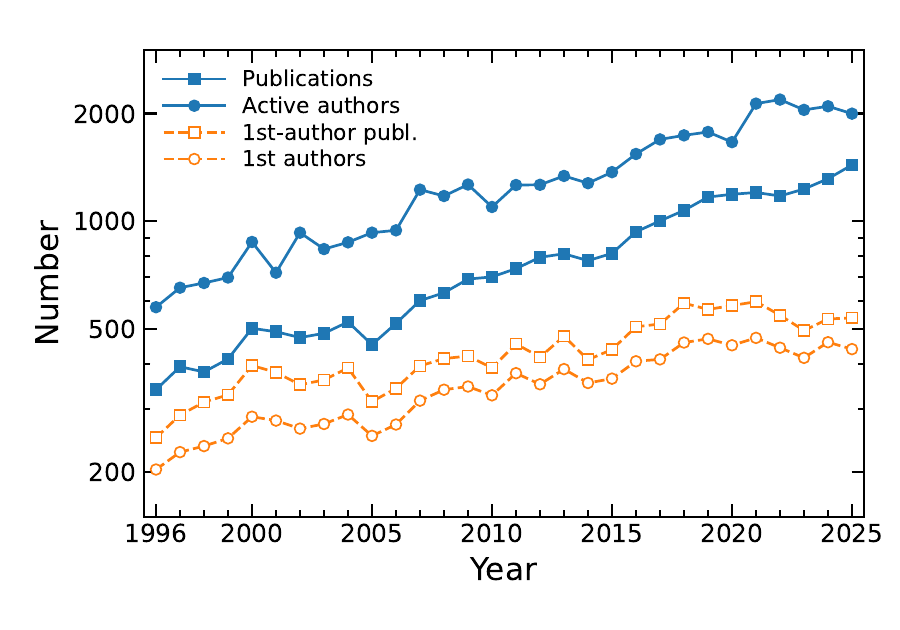}
 \end{center}
\caption{Annual community size and publication activity of the Japanese astronomy and astrophysics publication community from 1996 to 2025. Blue solid lines show overall activity: publications (filled squares) and active authors (filled circles). Orange dashed lines show the corresponding first-author activity: publications with a Japan-affiliated first author (open squares) and unique Japan-affiliated first authors (open circles). The vertical axis is logarithmic. {Alt text: Logarithmic line chart showing annual numbers of publications, active authors, publications with a Japan-affiliated first author, and Japan-affiliated first authors from 1996 to 2025. Overall publication and author activity increases more rapidly than the corresponding first-author activity, so that the relative contribution of first-author activity declines over the study period.}}
  \label{fig:community_size_and_publication_activity}
\end{figure}

Both the size of the active publication community and publication activity increased substantially during 1996--2025 (Figure~\ref{fig:community_size_and_publication_activity}). Community size is defined as the number of unique authors associated with at least one institution located in Japan at the time of publication who contributed to at least one target publication in a given year. Publication activity is represented by the annual number of target publications containing at least one such author. The number of active authors increased from 576 in 1996 to 1,998 in 2025, exceeding 2,000 in several years during the early 2020s. Over the same period, the annual number of publications increased from 339 to 1,437.

First-author activity also increased in absolute terms, but more slowly than the corresponding overall activity. The number of first authors increased from 203 in 1996 to 440 in 2025, while the number of publications with a Japan-affiliated first author increased from 249 to 538. Consequently, the fraction of active authors appearing as first authors decreased from 35\% to 22\%, and the fraction of publications with a Japan-affiliated first author decreased from 74\% to 37\%. The long-term expansion of the publication community was therefore increasingly associated with participation beyond first authorship.

\begin{figure}
 \begin{center}
  \includegraphics[width=90mm]{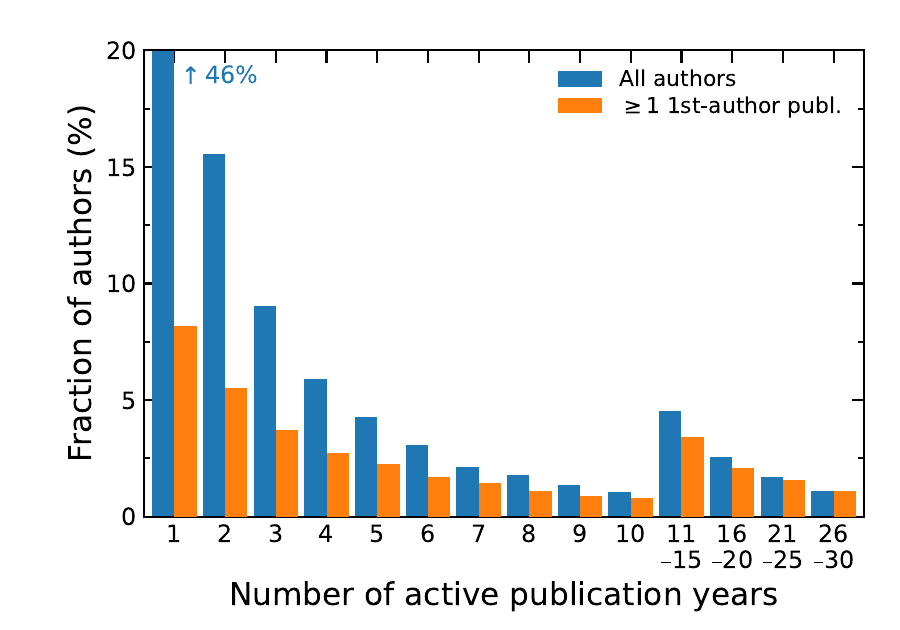}
 \end{center}
\caption{Distribution of observed active publication years per author during 1996--2025 and first-author experience within the study period. Blue bars show all authors in each participation-duration bin, and orange bars show the subset with at least one first-author publication. Both are expressed as percentages of all 9,910 unique authors. Multiple publications by the same author within a calendar year are counted once. Active publication years are shown individually for 1--10 years and grouped into 5-year bins thereafter. The vertical axis is limited to 20\% to show the distribution at longer participation durations more clearly; the 1-year bar for all authors exceeds this range and is annotated with its actual value. Because of the finite observation window, participation duration and first-author experience are both observation-window dependent and may be affected by boundary censoring. {Alt text: Grouped bar chart showing the percentage of authors by number of active publication years. The largest group consists of authors active in only one year and exceeds the plotted vertical range, with its actual value indicated on the figure. The number of authors declines strongly with increasing participation duration. Orange bars denote authors with at least one first-author publication and constitute an increasing fraction of the total toward longer participation durations.}}
  \label{fig:participation_duration_and_first_author_experience}
\end{figure}

Author participation varied substantially in duration. A total of 9,910 unique authors were identified over the full study period, of whom approximately 46\% were observed in only a single publication year (Figure~\ref{fig:participation_duration_and_first_author_experience}). The number of authors declined rapidly across the shorter observed participation durations, while a smaller but substantial group contributed over many years.

First-author experience was also strongly associated with observed participation duration. Among authors observed in only a small number of publication years, only a minority had at least one first-author publication during the study period. The fraction with first-author experience increased progressively toward longer participation durations. The distribution indicates both a large population of short-duration participants and a more persistent group in which first-author experience was increasingly common.

Because the observation window is limited to 1996--2025, the number of active publication years represents participation observed within the study period rather than complete publishing-career duration. Authors whose publication histories extend beyond either temporal boundary are therefore subject to left or right censoring. The first-author classification is similarly observation-window dependent. An author classified as having no first-author publication is one for whom no such publication was identified within the selected journals during 1996--2025, and may have had first-author publications before 1996, after 2025, or in journals outside the present dataset.

\subsection{Community renewal}\label{ssec:32}

\begin{figure}
 \begin{center}
  \includegraphics[width=90mm]{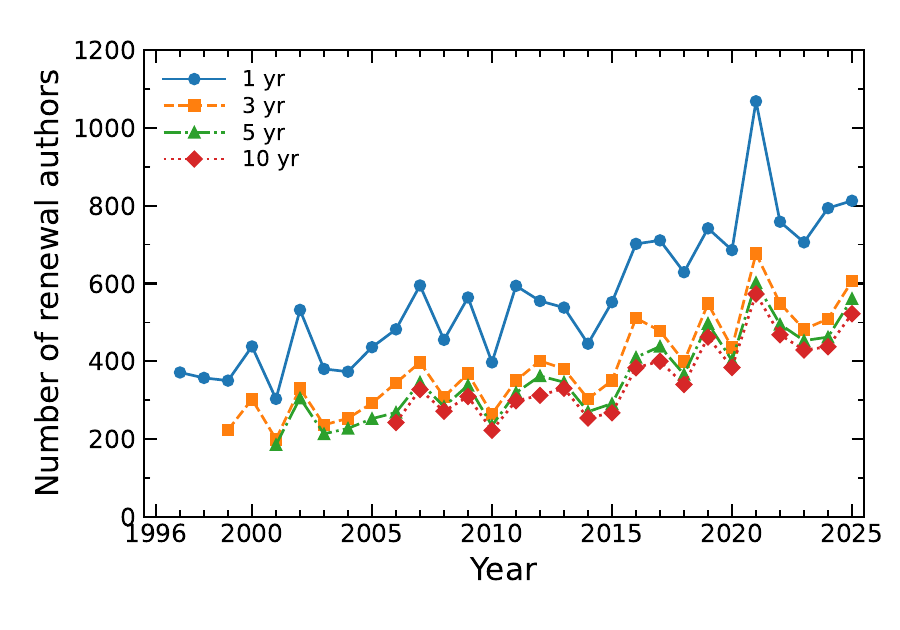}
 \end{center}
\caption{Temporal evolution of the number of renewal authors in the Japanese astronomy and astrophysics publication community for lookback windows of 1, 3, 5, and 10 years. Renewal authors in year $Y$ are defined as authors active in year $Y$ who were not observed during the preceding $T$ years, where $T$ denotes the lookback window; this definition includes both first-observed authors and authors returning after sufficiently long gaps. Each series begins only when a complete preceding lookback interval is available. Longer lookback windows generally produce smaller renewal-author counts, while the 5-year and 10-year results remain comparatively similar over their common time range. {Alt text: Four time-series curves show annual numbers of renewal authors for 1-, 3-, 5-, and 10-year lookback windows. Counts fluctuate from year to year but generally increase toward later years. Shorter lookback windows yield larger counts, while the 5-year and 10-year curves remain close over most of their overlapping period.}}
  \label{fig:renewal_author_counts}
\end{figure}

The number of renewal authors generally increased over the study period for all four lookback windows considered (Figure~\ref{fig:renewal_author_counts}), broadly reflecting the expansion of the active publication community. Renewal authors in year $Y$ are defined as authors active in that year who were not observed during the preceding $T$ years, and therefore include both authors first observed in year $Y$ and authors returning after an absence longer than the adopted lookback interval. To ensure a complete preceding interval, each series begins only after the corresponding lookback window is fully available: 1997, 1999, 2001, and 2006 for the 1-, 3-, 5-, and 10-year windows, respectively.

As expected from the definition, longer lookback windows generally classify fewer authors as renewal authors because any author observed during the longer preceding interval is excluded. The difference between the 5-year and 10-year windows was modest over most of their common time range. For example, the median renewal-author fraction was 25.8\% for the 5-year window and 23.6\% for the 10-year window. Extending the lookback interval from five to ten years therefore produced only a limited additional reduction in the identified renewal population. We consequently adopt the 5-year lookback window as the principal reference timescale in the following analysis.

\begin{figure}
 \begin{center}
  \includegraphics[width=90mm]{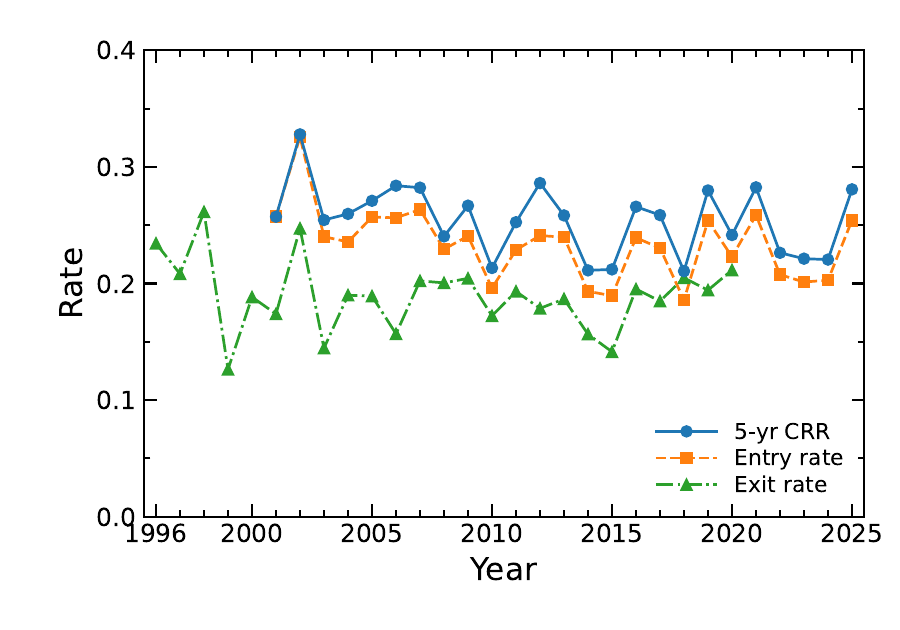}
 \end{center}
\caption{Temporal evolution of the 5-year Community Renewal Rate (CRR), entry rate, and exit rate in the Japanese astronomy and astrophysics publication community. The 5-year CRR is shown for 2001--2025, when a complete preceding 5-year lookback interval is available. The entry rate is shown only from 2001 onward to reduce boundary-censoring effects. The exit rate is shown only through 2020. The 5-year CRR includes both authors first observed in a given year and authors returning after publication gaps extending beyond the five-year lookback interval. {Alt text: Line chart showing the 5-year Community Renewal Rate and entry rate from 2001 to 2025 and the exit rate from 1996 to 2020. The CRR generally lies between about 21\% and 33\% and is consistently equal to or higher than the entry rate. The exit rate is generally lower than the CRR and entry rate over most of the period in which they can be compared.}}
\label{fig:crr_entry_exit}
\end{figure}

Using this reference timescale, the 5-year CRR fluctuated between approximately 21\% and 33\% during 2001--2025 (Figure~\ref{fig:crr_entry_exit}). In a typical year, approximately one quarter of the active publication community consisted of authors who had not been observed during the preceding five years. By construction, the 5-year CRR was equal to or higher than the entry rate because all authors first observed in year $Y$ were also absent from the preceding five-year interval. The CRR additionally included authors returning after publication gaps extending beyond the five-year lookback interval.

Across 2001--2025, these returning authors accounted for an average of 30.4 authors per year, corresponding to approximately 2.1 percentage points of the active community. Aggregated over the same period, returning authors represented 8.5\% of all renewal-author observations. The displayed entry rate ranged from 18.6\% to 32.6\%, with a median of 23.9\%. The exit rate, shown only through 2020 to reduce right-boundary censoring, ranged from 12.6\% to 26.2\%, with a median of 19.0\%. Over most of the uncensored comparison period, the exit rate was lower than both the 5-year CRR and the entry rate, although this ordering is an empirical feature of the present data rather than a consequence of the definitions.

\section{Discussion}\label{sec:4}

The Japanese astronomy and astrophysics publication community expanded substantially during 1996--2025 while exhibiting considerable heterogeneity in participation patterns. Although many authors were observed in only a single publication year within the study period, a substantial population of authors with long observed participation periods was also identified. For context, the annual active-author population was approximately 0.2\% and 0.5\% of the head-count researcher populations reported for Japan's universities and national-level research institutions\footnote{National research institutes, special corporations, and independent administrative institutions.} in 1996 and 2024, respectively \citep{NISTEP2025}. These are only order-of-magnitude comparisons because the two datasets differ in disciplinary coverage and researcher definitions.

The expansion of the publication community was also accompanied by a change in author-position patterns. Although first-author activity increased in absolute terms, it grew more slowly than overall participation. The fraction of active authors appearing as first authors decreased from 35\% in 1996 to 22\% in 2025, while the fraction of publications with a Japan-affiliated first author declined from 74\% to 37\%. At the same time, first-author experience became progressively more common among authors with longer observed participation histories. These patterns indicate that the expansion of the community increasingly involved participation beyond the first-listed author position, while sustained participation was more often associated with at least some first-author experience. This pattern is consistent with an increasing role of broader collaborative participation, including large and international collaborations, although the present analysis does not directly identify the mechanisms responsible for the change.

The analysis of community renewal suggests that the publication community maintained a persistent level of demographic renewal from 2001 onward, the period for which the 5-year CRR could be evaluated. Most renewal authors were first observed participants. Authors returning after publication gaps extending beyond the five-year lookback interval represented 8.5\% of renewal-author observations aggregated over 2001--2025. The CRR therefore largely tracked entry while retaining a distinct returning-author component, providing a community-level measure of renewed active membership rather than first-time entry alone once a complete lookback interval was available. Combined with the presence of authors with long observed participation periods, these patterns indicate that the community sustained continuity while undergoing recurring changes in its active membership.

This coexistence of continuity and renewal is broadly consistent with the institutional evolution of Japanese astronomy and astrophysics identified in a separate bibliometric analysis \citep{Fujiwara2026b}. That study found that the number and diversity of participating Japanese institutions increased during 1996--2025, while publication activity remained concentrated around a persistent core of leading institutions. Although the two studies use different publication-selection frameworks and do not establish a direct causal relationship between author turnover and institutional broadening, their results together suggest that the expansion of the publication community occurred through changes in both its individual membership and its institutional base, while retaining persistent core components.

A simple order-of-magnitude estimate suggests that the influx of graduate students alone is unlikely to explain the observed renewal. Approximately 55 individuals completed or left doctoral programs in astronomy and astrophysics annually during 2009--2017 \citep{Okumura2021}. Assuming approximate steady state, a doctoral intake of about 50 per year, a master's-to-doctoral enrollment ratio of 2.5--3.0:1, somewhat higher than the national ratio reported in the School Basic Survey \citep{MEXTSchoolBasicSurvey}, and two- and three-year enrollment periods imply an annual master's intake of order 200 and a total graduate-student population of roughly 550. This estimate is broadly consistent with the 517 and 673 student members reported by the Astronomical Society of Japan\footnote{$\langle$https://www.asj.or.jp/jp/about/outline/$\rangle$.} in 2020 and 2025, respectively. An annual master's intake of approximately 200 corresponds to about 40\% of the roughly 500 renewal authors observed annually in recent years. Even if all newly enrolled master's students appeared as first-time authors, they would account for less than half of the renewal-author population, and the actual fraction is likely smaller. The remainder may reflect returning authors, international or institutional mobility, short-term affiliations, entry from adjacent fields, and participation in large collaborations.

Community renewal should not be interpreted as evidence of stable career conditions or workforce sustainability. A survey of astronomy and astrophysics PhD holders in Japan found that nearly 80\% entered fixed-term positions immediately after receiving their doctorate, 35\% had never held a permanent position at the time of the survey, and the transition to permanent employment was substantially slower for post-2000 cohorts than for earlier cohorts \citep{Okumura2021}. Such career conditions may contribute to author turnover, but the present bibliometric design cannot determine employment status or associate observed entry and exit with specific career transitions.

This study has several limitations. The analysis is based on articles, reviews, and letters published in the selected journal set and indexed by Scopus, and therefore characterizes the observed publication community rather than the complete population of astronomy and astrophysics researchers associated with Japanese institutions. Researchers publishing exclusively outside the selected journal set are not represented in the dataset. Longitudinal author identification depends on the Scopus Author ID, while the classification of Japan-affiliated authors depends on the institutional affiliations recorded at the time of publication. Potential splitting or merging of Scopus Author IDs may introduce uncertainty into individual participation histories and, consequently, into the derived entry, exit, and renewal classifications. The distribution of observed participation duration is also affected by the finite observation window, particularly for authors whose publication records extend to the beginning or end of the study period, and should not be interpreted as a distribution of complete publishing-career duration. In addition, the entry and exit rates should be interpreted with caution near the boundaries of the study period because of left- and right-censoring, respectively.

\section{Conclusion}\label{sec:5}

Using Scopus and SciVal data, we characterized the demographic evolution of the Japanese astronomy and astrophysics publication community during 1996--2025. The active community expanded from approximately 600 authors in the late 1990s to more than 2,000 by the early 2020s, while participation remained highly heterogeneous: 46\% of authors were observed in only one publication year, but a smaller core had long observed participation periods.

The 5-year CRR typically remained near one quarter of the active community, ranging from approximately 21\% to 33\% during 2001--2025, indicating persistent renewal alongside community growth. An order-of-magnitude comparison suggests that newly entering graduate students alone are unlikely to account for the full renewal-author population, which likely reflects multiple pathways including first-time participation, return after publication gaps, mobility, short-term affiliations, entry from adjacent fields, and large collaborations. This framework provides a basis for comparative analyses of scientific publication communities.

\begin{ack}
This work was supported by JSPS KAKENHI Grant Number JP26K03645. The author thanks Elsevier Ltd.\ for increasing the Scopus API access limits to facilitate the data collection for this project. The author also thanks the anonymous referee of \citet{Fujiwara2026b} for highlighting during the review of that paper the importance of bibliometric examination of first-author structure for understanding research activity in the Japanese astronomy and astrophysics community.
\end{ack}

\end{document}